\documentclass[10pt,showpacs,amsmath,amssymb,floatfix,superscriptaddress,longbibliography,column,eqsecnum]{revtex4-2}
\usepackage{booktabs}
\usepackage{graphicx}
\usepackage{stfloats}
\usepackage{balance}
\usepackage{xcolor}
\usepackage[verbose]{hyperref}
\hypersetup{colorlinks=false}
\usepackage{tabularx}
\usepackage{ifthen}
 
\begin{document}

\title{The convolutional neural networks for analysing the micro-cavity array multi-mode quantum frequency comb spectrum features}

\author{Hang Shen}
\address{Department of Physics,College of Science,Hangzhou Dianzi University,Zhejiang 310018,China.}

\author{Chaoying Zhao*}
\address{Department of Physics,College of Science,Hangzhou Dianzi University,Zhejiang 310018,China.}
\address{State Key Laboratory of Quantum Optics Technologies and Devices, Shanxi University,Taiyuan,030006,China}
\address{Zhejiang Key Laboratory of Quantum State Control and Optical Field Manipulation, Hangzhou Dianzi University, Hangzhou 310018,China}
\email[*]{zchy49@163.com}

\begin{abstract}
\textbf{Micro-ring play an important role in light–matter interaction. The nonlinear Kerr micro-ring generated multi-mode optical frequency combs (OFCs) can lock any frequency at any time. Based on the quantum cavity electrodynamics, we have gave out the quantum comb tooth formation principle of multi-mode OFCs of Kerr micro-ring. In this work, we furthermore put forward a novel spectrum characteristics recognition method based on the quantum micro comb. we use machine learning to automatically extract quantum OFCs spectrum characteristics based on the micro-ring array and wave-guide array coupling systems. Our quantum theorem analysis is in good agreement with numerical simulation results. The average accuracy of single parameter identification attains to $99.5\%$, and the average accuracy of two parameters identification attains to $97.0\%$. We can quickly distinguish the quantum OFCs spectrum characteristics by multi-parameter changes. Our method has a potential to improve the multi-mode recognition accuracy.}  
\end{abstract}

\maketitle

\textbf{1. Introduction}\\

Micro-cavity~\cite{jiang2020whispering,xue2019ultrasensitive}can detect the information contained in the transmission spectrum~\cite{kuhnline2013interfacing,chen2023optical,wade2015rapid}. The optical spectrum information can be extracted from the spectrum shift of the whispering gallery mode (WGM) formed by the micro-cavity. There are many different kinds of optical extraction devices~\cite{gohring2010detection,chen2023optical,li2021smart}. Ultrahigh quality factor and small mode volume WGM micro-cavity is difficult to achieve a single WGM light field. The multi-mode WGM light field is limited in practical applications due to the lack of a mode selection strategy. Considering the micro-ring is very suitable for integrated systems due to their small size, low sample consumption, and multiplexing~\cite{wade2016applications,sun2011optical,gohring2010detection,li2021smart}, we assume that we can adopt a micro-ring array to realize mode selection of the resonance spectrum. The extraction process of a conventional information detection method~\cite{nath2007applications} is usually cumbersome, laborious, and time-consuming. In addition, not all components can be extracted at once. In order to identify all components, we use multiple linear regression, principal component regression, and partial least squares algorithms to decompose the spectrum into single component~\cite{wehrens2007pls}. If we look at the spectrum as a linear superposition of each individual component, it will lead to a large number of indistinguishable overlapping spectra. Researchers have proposed some models to distinguish these overlapping spectra~\cite{roggo2007review}. Some popular spectral information analysis methods require a large database that can contain the spectra of each individual component, which include ultraviolet spectrophotometry, Raman spectroscopy, nuclear magnetic resonance spectroscopy, and so on~\cite{toumi2014review}. Machine learning has a unique learning ability, which can build data-driven models to identify features from the database by itself. In recent years, deep learning~\cite{lecun2015deep,schmidhuber2015deep,robison2019precision,vamathevan2019applications,cheng2014machine} has received extensive attention, such as helping researchers optimize the design of micro-ring based on surface plasmon resonance~\cite{moraru2010using,zhao2008machine}.

 We have realized the quantum optical frequency combs (OFCs) in a high-quality micro-ring resonator through continuous wave pumping~\cite{shen2023method}. The multi-mode quantum OFCs light source can make spectral information detection~\cite{li2021smart}. Due to the different components affecting the quantum OFCs spectrum of the micro-ring resonator. Combined with our previous quantum theoretical work on the comb tooth formation principle of quantum OFCs~\cite{shen2023method}, we find that machine learning can effectively distinguish various micro-ring parameters, and these parameters' changes theoretically will affect the quantum OFCs spectrum. Therefore, we can extract spectrum information by using quantum OFCs and analyze the quantum OFCs spectrum information by using machine learning. \\

\begin{figure}
\includegraphics[width=0.48\textwidth]{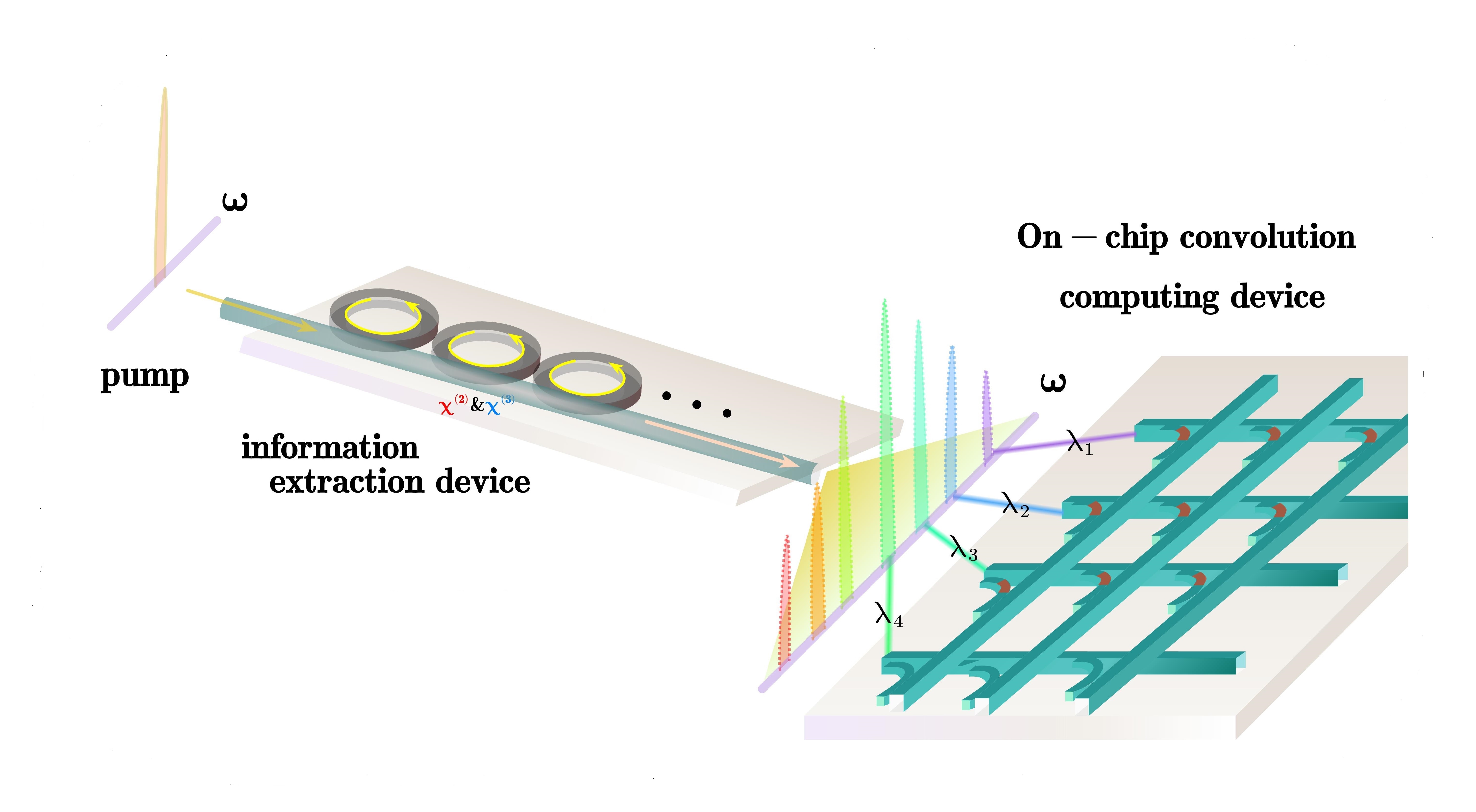}
\caption{\label{fig:1} The information extraction device is pumped by continuous laser. The different comb teeth are multiplied by on-chip convolution computing device.}
\end{figure}

\textbf{2. Quantum OFCs spectrum information extraction}\\

The micro-ring resonator array made of nonlinear materials has a high $Q$ factor and can realize the accumulation of optical energy as shown in Figure 1. Moreover, materials have some characteristics, such as piezoelectric, magneto-electric, and thermo-electric. The low pump power limits the comb teeth number of quantum OFCs, and while the high pump power causes the micro-ring thermal effects and interferes with the detection result. The Kerr micro-comb results from the four-wave mixing process~\cite{chen2023optical}, and an appropriate micro-ring array doesn't destroy the material composition. In Figure 2(a), Kerr micro-comb~\cite{guidry2022quantum} is only related to the third-order nonlinear effect. Considering the micro-ring has a stronger second-order nonlinear effect, the second-order nonlinear effect of the Kerr micro-comb can be controlled by changing the size of the micro-ring array during the manufacturing process~\cite{guo2016second}. In Figure 2(b), the second-order and the third-order nonlinear effects make the Kerr micro-comb~\cite{bruch2021pockels} have more comb teeth. Therefore, we can extract more spectrum information theoretically. In fact, the second-order nonlinear effects being far less than third-order nonlinear effects, in our theoretical work, we only consider the third-order nonlinear effect. We assume that the manufacturing process of each micro-ring is the same.

\begin{figure}
\centerline{\includegraphics[width=0.48\textwidth]{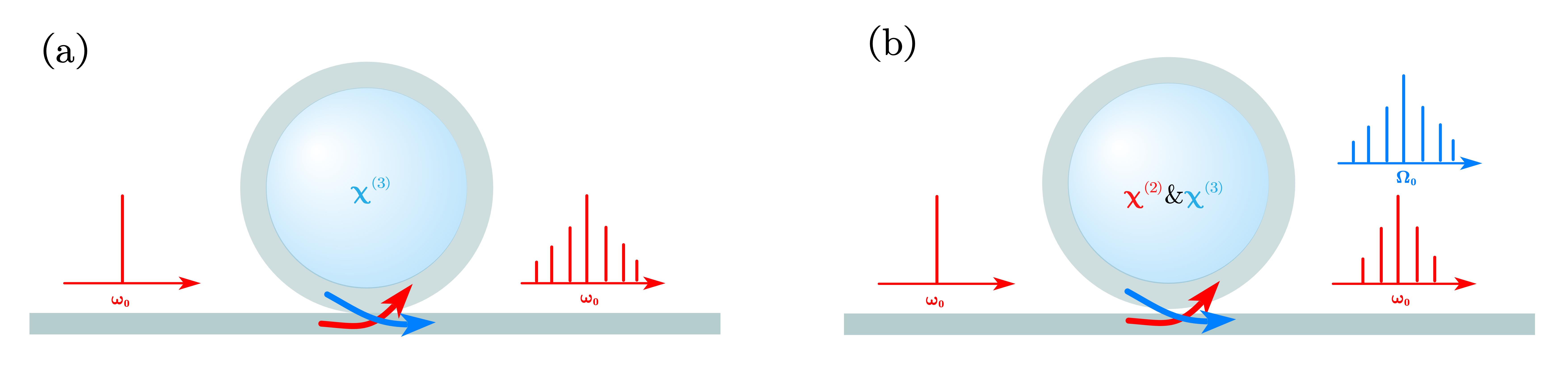}}
\caption{\label{fig:2} Kerr micro-comb is controlled by (a) third-order nonlinear effect. (b) second-order and third-order nonlinear effect. The blue line corresponds to second harmonic mode and the red line corresponds to fundamental mode.}
\end{figure}

When the environmental change inside the micro-ring array leads to the change of the WGM light field of the micro-ring array inner wall. The different structural parameters of the micro-ring array~\cite{li2021smart} will affect the quantum OFCs spectrum. In order to explain how different micro-cavity structural parameters affect the quantum OFCs spectrum, we need to establish the relationship between the quantum OFCs spectrum and the structural parameters of the micro-ring array. Based on the Hamiltonian of quantum OFCs in the micro-ring array, we give out the quantum OFCs dynamic equation. We can understand its dynamic process phenomenologically by simulating the quantum OFCs spectrum of the micro-ring array.

The Hamiltonian of the mode family in the micro-ring array is

\begin{equation}
\hat{\mathcal{H}}_{0}=\sum_l{\hbar\omega _{a_{l}}\hat{a}_{l}^{\dagger}\hat{a}_{l}},
\end{equation}

where $l$ represents different photons in the frequency domain. $\hat{a}_{l}$ represents the photon annihilation operator, $\hat{a}_{l}^{\dagger}$ represents the photon creation operator, $\omega_{a_{l}}$ represents the photon angular frequency.  

In the following simulation, after the pump light continuously inputs into a high $Q$ factor micro-ring array, for the convenience of calculation, we set the material of all micro-rings is AlN. $R$ is the radius of each micro-ring. When the energy reach a certain threshold, and a strong nonlinear optical process occur. We only consider the third-order nonlinear effect.

\begin{equation}
\hat{\mathcal{H}}_{\chi^{\left(3 \right)}}=\sum_{klmn}{\hbar g_{klmn}\hat{a}_{k}^{\dagger}\hat{a}_{l}^{\dagger}\hat{a}_{m}\hat{a}_{n}},
\end{equation}

\begin{equation}
\hbar{g_{3}}\approx \sqrt{\frac{\left(\hbar\omega _1\right)^2\hbar\omega_{2}\hbar\omega_{3}}{\left(\epsilon_{0}\epsilon_{2}\right)^2\epsilon_{1}\epsilon_{3}}}\frac{1}{2\pi R}\frac{1}{A_{eff}}\chi^3\xi,
\end{equation}

where the third-order nonlinear coefficient $g_{klmn}$ represents the intensity of the four-wave mixing process. $A_{eff}$ is the effective mode area on the cross section. $\xi$ is the mode overlap factor on the cross section. We have $0\leqslant\xi\leqslant 1$. $\epsilon_{i}$ is the relative dielectric constant at the frequency of $\omega_{i}$.
When the external pumping in $p$ mode, we have

\begin{equation}
\mathcal{H}_{pump}=\hbar\epsilon_{a_{p}}\left( \hat{a}_{p}e^{i\left(\omega_{p}+\delta_{p}\right)t}+\hat{a}_{p}^{\dagger}e^{-i\left(\omega_{p}+\delta_{p} \right)t}\right),
\end{equation}

where  
 
 \begin{equation}
 \epsilon_{p}=\sqrt{\frac{kP_{in}}{\hbar\omega_{p}+\delta_{p}}},
 \end{equation}

where $k={\omega_p}/{2Q}$ represents the internal and external coupling loss. $\delta_{p}$ represents the pump detuning. 

In order to get rid of the dependence of the micro-ring array on time, we adopt the rotating coordinate; the unitary transformation matrices are

\begin{equation}
\hat{U}\left(t\right)=e^{i\hat{R}t},
\end{equation}

the mode family frequency is  

 \begin{equation}
 \omega_{a}=\omega_{0}+d_{1}l+\delta,
 \end{equation}
 
where $l$ is the interval between the optical light frequency, $d_{1}$ is the center frequency of the mode family that can be obtained from $2\pi FSR$, the free spectral range $FSR={c}/{2\pi n_{eff}R}$, $n_{eff}$ is the effective refractive index of the medium, and $c$ is the speed of light in vacuum. 

\begin{equation}
\hat{R}=\sum_l{\left(\omega_{p}+d_{1}l+\delta \right)}\hat{a}_{l}^{\dagger}\hat{a}_{l},
\end{equation}

 Ignoring the higher-order dispersion, the eigen-frequency can be expressed by 

 \begin{equation}
 \omega_{j}=\omega_{0}+d_{1}l+\frac{d_{2}l^2}{2},
 \end{equation}
 
 where $\beta_{2}$ is the second-order group velocity dispersion,
 
 \begin{equation}
 d_{2}=-\frac{c}{n_{eff}}d_{1}^{2}\beta_{2},
 \end{equation}
 
The detuning of the mode family can be expressed by 

\begin{equation}
\delta_{a_{l}}=d_{2}l^2-\delta. 
 \end{equation}
 
Combining Eqs.(1) and (2), the system Hamiltonian becomes 

\begin{equation}
\hat{\mathcal{H}}_{sys}=\sum_l{\hbar\delta _{a_{l}}\hat{a}_{l}^{\dagger}\hat{a}_{l}}+\hat{\mathcal{H}}_{\chi^{\left(3\right)}}+\hbar\epsilon_{a_{0}}\left(\hat{a}_{0}+{\hat{a}_{0}}^{\dagger}\right),
\end{equation}

We decompose the optical state into a mean field solution (assuming a coherent state) and quantum fluctuations.

\begin{equation}
\hat{a}_{l}\left(t\right) =\alpha_{l}\left(t\right) +\hat{a}_{l}\left(t\right). 
\end{equation}

We remove the first-order term and constant term, only keep the second-order term, and ignore the higher-order term, the Heisenberg's equation of motion can be expressed by

\begin{equation}
\dot{\hat{X}}=-i\left[X,\mathcal{H}\right]-k_{x}X,
\end{equation}

Based on the cavity quantum electrodynamics theory, we can describe the dynamic equation of quantum OFCs in the following form

\begin{equation}
\frac{d}{dt}a_{l}=\left(-i\delta_{a_{l}}-\kappa \right)a_{l}-i\sum_{k,l,m}{2}g_{3}a_{k}^{\dagger}a_{l}a_{m}-i\epsilon_{p}\delta_{p}\left(l-p\right),
\end{equation}

where $\kappa$ is the total coupling loss.\\

\begin{figure}
\includegraphics[width=0.48\textwidth]{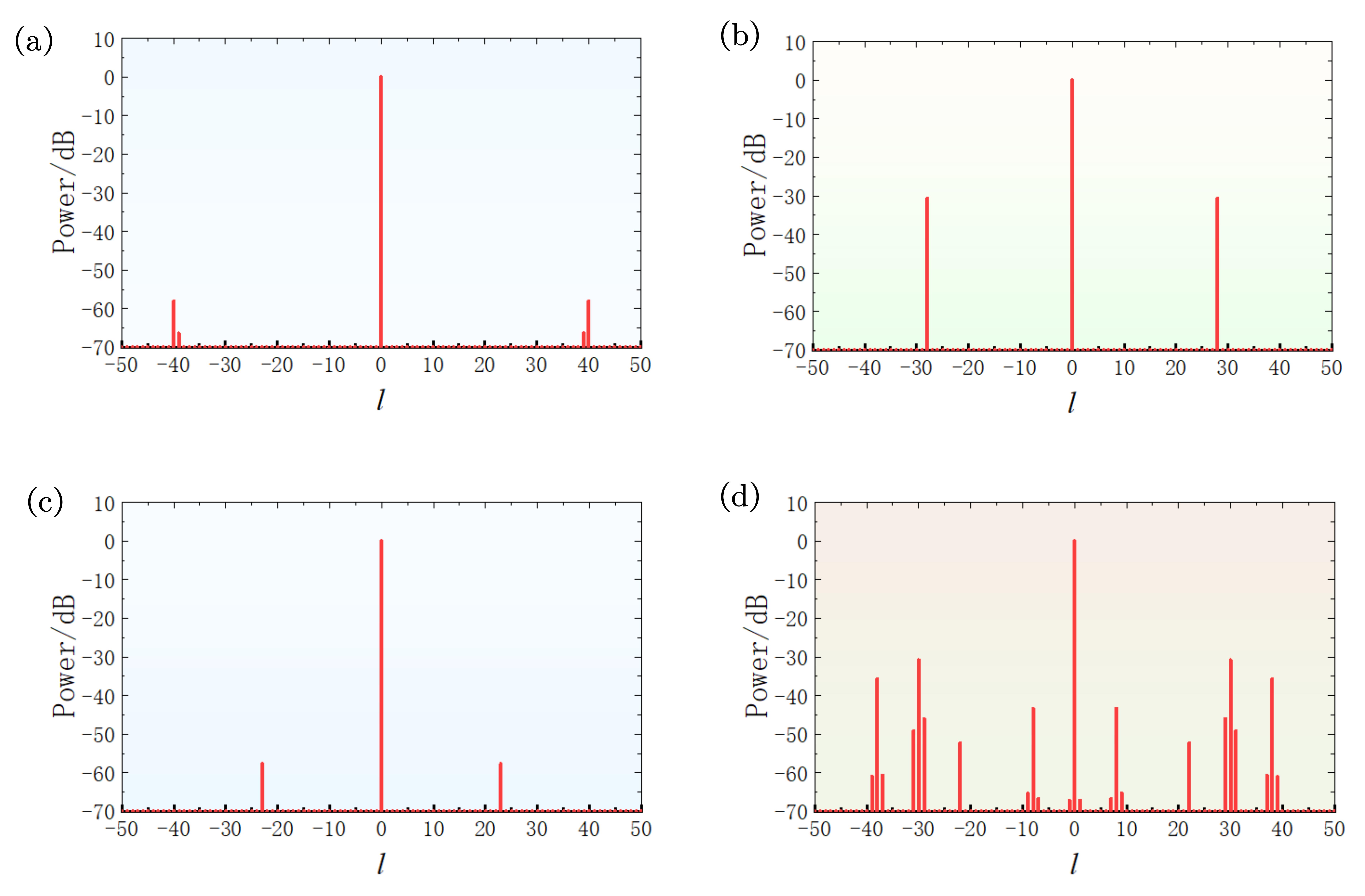}
\caption{\label{fig:3} The threshold power of different modes $l$ in the fundamental
frequency modes. The parameters used in the simulation are $k=1$, $\varepsilon_{p}=4.2\sqrt{2}$. (a) $d_{2}=0.003$, $g_{3}=1$; (b)$d_{2}=0.004$,  $g_{3}=1$; (c) $d_{2}=0.005$, $g_{3}=1$; (d) $d_{2}=0.004$, $g_{3}=1.4$.}
\end{figure}

We adopt Fourier and inverse Fourier transform~\cite{hansson2014numerical}, and use the fourth-order Runge-Kutta method to simulate the quantum OFCs spectrum of $128$ modes in the micro-ring array. In order to facilitate the data processing, we take the normalized dimensionless parameters~\cite{chembo2010modal}: $d_{2}=0.004$, $k=1$, $g_{3}=1$, and $\varepsilon_{p}=4.2\sqrt{2}$. As shown in Figure 3, we can clearly see that the influence of $g_{3}$ and $d_{2}$ on the quantum OFCs spectrum is different, which is consistent with the formation position of the main comb teeth~\cite{shen2023method}. With the increase of $d_{2}$, from Figures 3(a) and 3(c) we can see that the lateral peak will be closer to the center of the spectrum, the speed of the lateral peak approaches to the central mode of the spectrum; it isn't a simple linear relationship~\cite{shen2023method}, but a complex nonlinear relationship. This is beyond the scope of this article's discussion, and we will investigate this issue in our future work. With the increasing of $g_{3}$, we find that the position of the side peak does not change obviously, as shown in Figures 3(b) and 3(d),  there are more comb teeth appearing in the spectrum. Our primary consideration next is how to select the spectrum feature extraction method. As shown in Figure 4, we think that the machine learning method is undoubtedly the best choice at present; we can use it to extract data features independently from training data.\\

\textbf{3. Quantum OFCs spectrum information processing}\\

\begin{figure}
\includegraphics[width=0.48\textwidth]{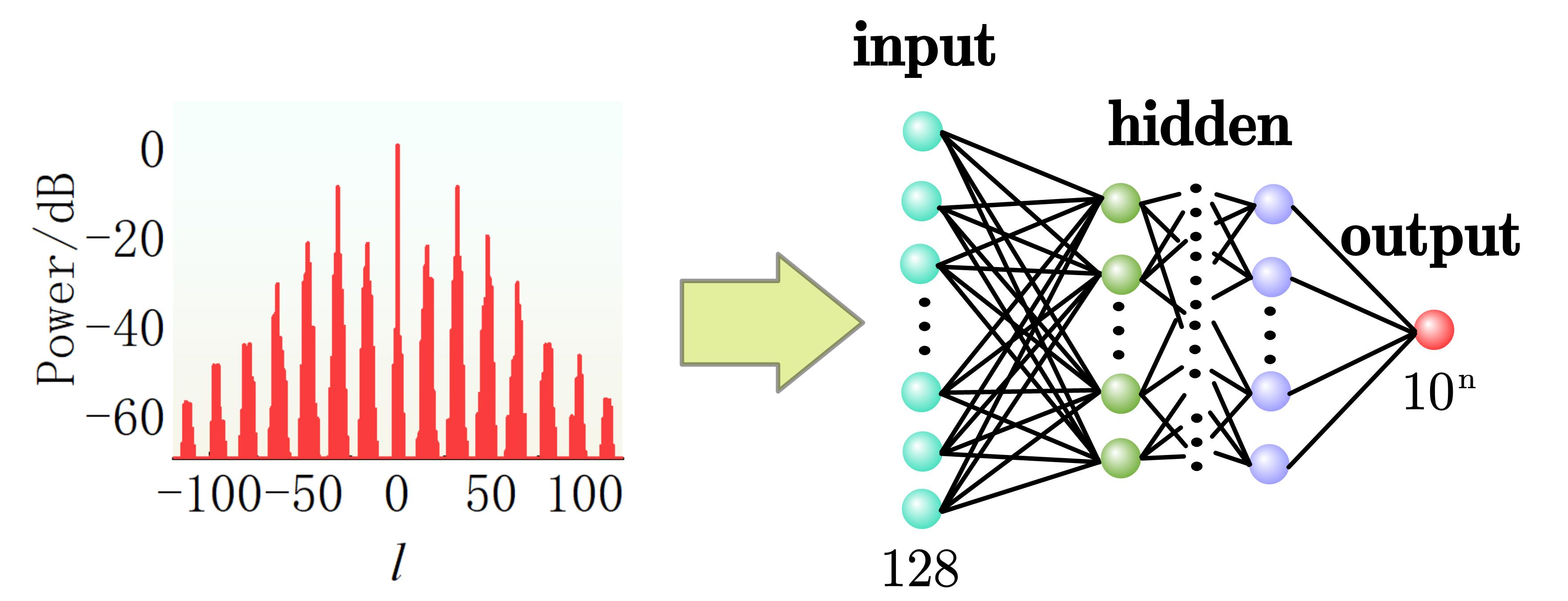}
\caption{\label{fig:4} Simple schematic diagram of substituting optical frequency comb into neural network calculation. The input image is the threshold power of different modes $l$ in the fundamental frequency modes. A simple network structure including one input layers, two hidden layers, and one output layer. The initial convolution layer is used for preliminary feature extraction. The convolution kernel is a tensor matrix. The basic features of the image are captured to lay the foundation for subsequent feature extraction.}
\end{figure}

 We choose the light intensity as the measurement quantity in the photonic integrated device. Figure 5(a) shows the coupled unit structure between horizontal waveguides and vertical waveguides~\cite{guo201670}. The waveguide coupling array structure can realize linear activation as shown in Figure 5(b). There are two waveguide arrays in horizontal and vertical directions. We control the coupling ratio between the horizontal waveguide and the vertical waveguide to obtain the optimal light intensity.
\begin{equation}
I_{out}=k_{11}I_{1}+k_{21}I_{2}+k_{31}I_{3}+k_{41}I_{4},
\end {equation}

 Next, we analyze the concrete calculation mechanism in the waveguide array. The number of horizontal waveguides corresponds to the optical modes number of quantum OFCs spectrum for the micro-ring array, and different optical modes correspond to different angular frequencies. A picture can be divided according to pixel points. As shown in Figure 5(c), the $16$ light modes of the quantum OFCs spectrum correspond to $16$ parts. Figure 5(d) shows the photon calculation process of convolution activation. A series of optical modes on the left side of each row waveguide are multiplexed by a wavelength division multiplexer, and the multiplexed optical modes will perform the product operation of the same convolution kernel elements. The function of the left wavelength division multiplexer is to recombine and group the optical modes to complete the convolution operation in the same waveguide, and present the results of the convolution operation. The linear activation and convolution activation can help us realize machine learning, the key point is the multiplication calculation at the real photon level. In order to realize the multiplication calculation, each column of the waveguide array represents the calculation of a convolution kernel, the number of linear neurons can be freely controlled by selecting the number of vertical waveguides. In simulation, we set different convolution kernel elements by taking the coupling ratio between each column and the horizontal waveguide. It is possible to realize the calculation of the neural network on the optical integrated device by different combinations of Figure 5(b) and 5(d). The four layers were fully connected and activated by rectified linear units (ReLU) function. The training model was set up by using TensorFlow in Figure 5(e). A four-layer perceptron neural network (one input layer, two hidden layers, and one output layer) was adopted to implement a regressor to predict (1) $d_{2}$: the input layer has $128$ neurons, the two hidden layers have $64$ and $36$ neurons, and the output layer has $10$ neurons. (2) $g_{3}$: the input layer has $128$ neurons, the two hidden layers have $1024$ and $60$ neurons, and the output layer has $10$ neurons.(3) $d_{2}$ and $g_{3}$: the input layer has $128$ neurons, the two hidden layers have $1024$ and $256$ neurons, and the output layer has $100$ neurons.\\

\begin{figure}
\includegraphics[width=0.48\textwidth]{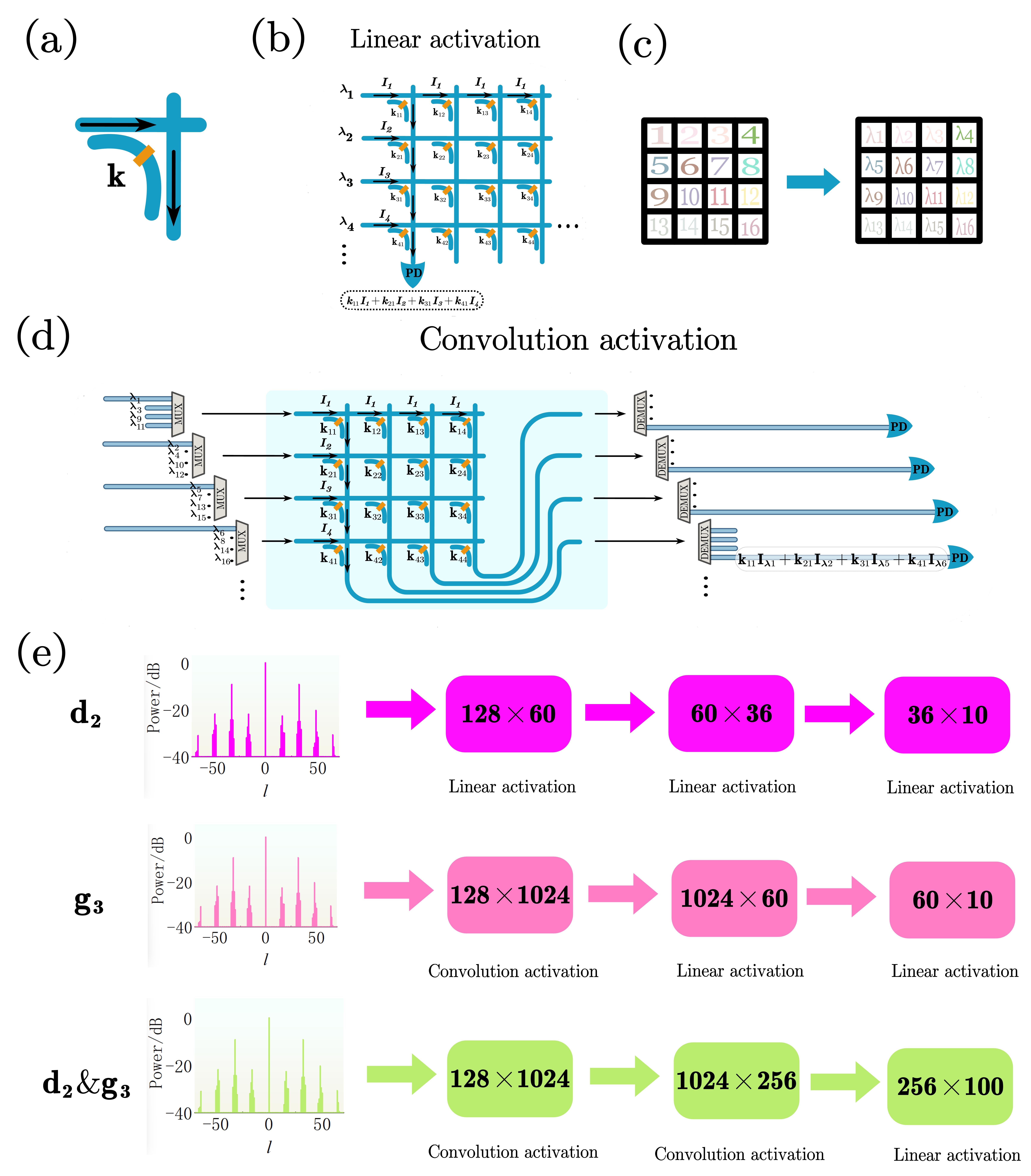}
\caption{\label{fig:5} Integrated theoretical method for detecting quantum OFCs spectrum characteristics. (a) The coupling device between the horizontal waveguide and the vertical waveguide can change the waveguide gap by changing the temperature result in the coupling rate change. (b) Schematic diagram of optical integrated device for linear activation. (c) The different frequencies of the input spectrum are regarded as different pixel points of a picture for convolution calculation. (d) Each wavelength can carry independent information and then be integrated through wavelength division multiplexing (WDM) for corresponding feature processing. Afterwards, the wavelengths carrying information are separated through WDM to achieve effective extraction of parallel information. (e) The structure of neural networks for identifying threshold power spectrum with different parameters. The tangerine corresponds to the impact of single $d_{2}$ parameter changes, the pink corresponds to the impact of single $g_{3}$ parameter changes, and the green corresponds to the impact of $d_{2}$ and $g_{3}$ two parameters change.}

\end{figure}
\begin{figure}
\includegraphics[width=0.48\textwidth]{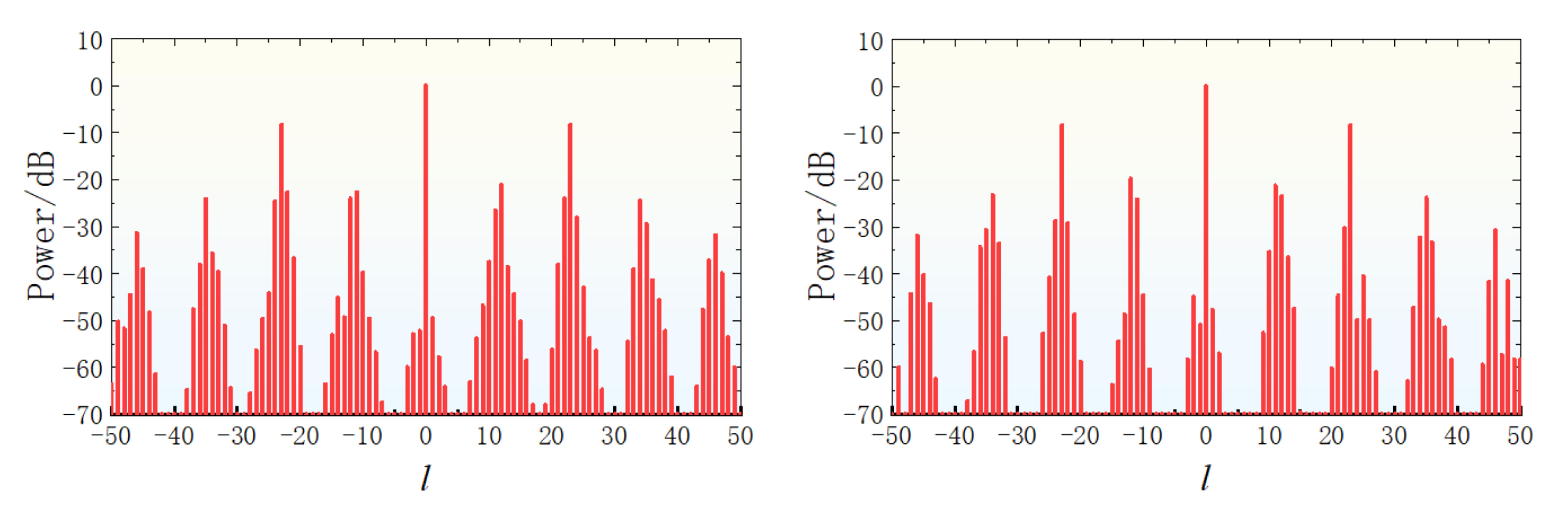}
\caption{\label{fig:6} Spectral differences caused by the threshold power of different modes $l$ in the initial state of the micro-ring  array is randomly taken as a different random phase.}
\end{figure}

\textbf{4. Feasibility and accuracy analysis}\\

In order to illustrate the feasibility and superiority of our theoretical model, we furthermore make the following simulation calculation. First of all, machine learning can be used to extract and classify the quantum OFCs spectrum characteristics of micro-ring array; we use training data to train neuron parameters. In the second section, the quantum OFCs spectrum characteristics change with the micro-ring structural parameters. The initial parameters: $d_{2}=0.004$, $k=1$, $g_{3}=1$, $\varepsilon_{p}=4.2\sqrt{2}$. The evolution of the light field in the micro-ring takes the same time, and different spectra and intensities of different light modes in the spectrum can be obtained by continuously changing the structural parameters of the micro-ring. We adjust the size of $d_{2}$ from $0.002$ to $0.006$, with the size interval of $0.0004$, and get $10$ groups of quantum OFCs spectrum data. The corresponding light intensity of each mode can be represented by $\mathrm{dB}$, and the $10$ data are labeled from $0$ to $9$ so as to distinguish different $d_{2}$. The initial value of $a_{l}$ is taken as a random small quantity in complex form, which represents the noise term. We set a random phase distribution value between $-\pi$ and $\pi$, and then use the random phase to calculate the noise power spectrum. We set different cycle times and obtain the different samples, as shown in Figure 6. We can observe the frequency spectrum very clearly at any time. As a result, the spectrum has a slightly different character when there are a small amount of changes in the initial spectrum, but the overall spectrum features are basically similar. The different samples can be obtained by increasing the number of spectrum simulations. By adjusting the size of $g_{3}$ from $1$ to $1.5$ with the size interval of $0.05$, we obtain $10$ groups of quantum OFCs spectrum data, and by simulating the spectrum for different times, we furthermore obtain the same type of data samples.

\begin{figure}
\includegraphics[width=0.48\textwidth]{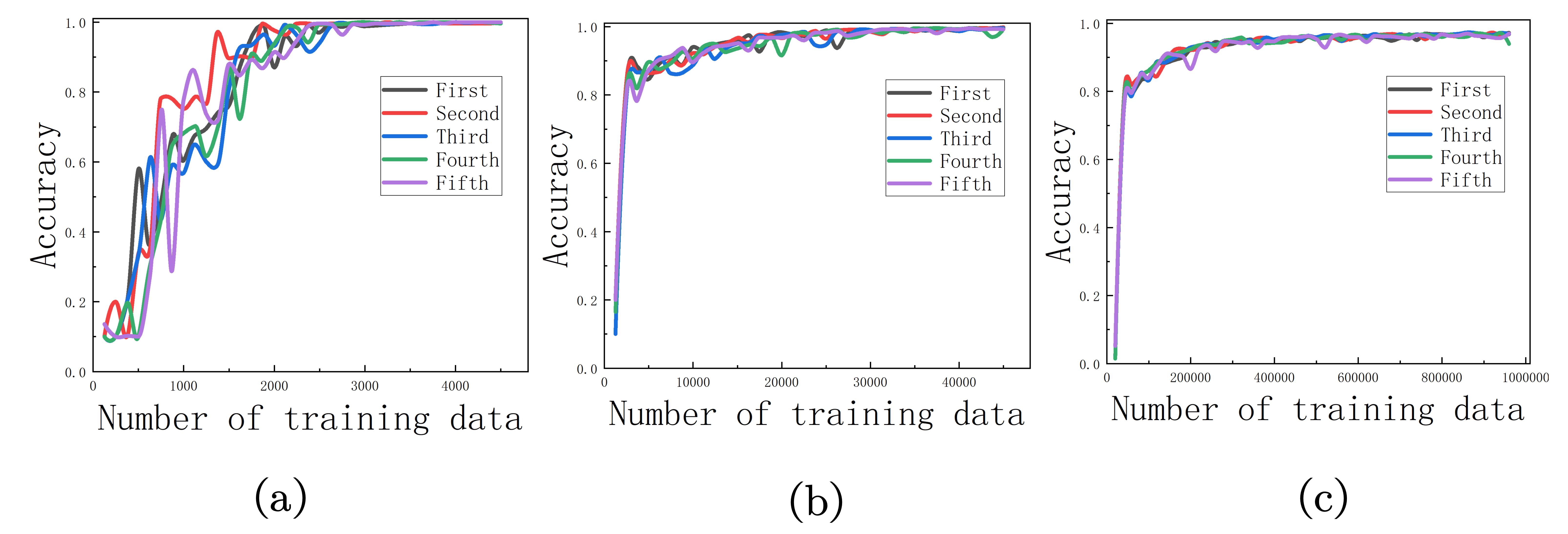}
\caption{\label{fig:7} Visualization of the recognition accuracy with the increasing of training data quantity under different parameter changes (five rounds machine learning process, the five colors in the small box represent the rounds). (a) $d_{2}$ parameter (detailed data see Table 1). (b) $g_{3}$ parameter (detailed data see Table 2). (c) $d_{2}$ and $g_{3}$ parameters (detailed data see Table 3).}
\end{figure}

Due to $d_{2}$ and $g_{3}$ having different effects on the quantum OFCs spectrum, we use the computer to train the parameters of the neural network through Pytorch (the neural network setup as shown in Figure 5(d)). We use $300$ samples to measure the $d_{2}$ parameter change. The accuracy of is $99.8\%$ as shown in Figure 7(a), and the calculation formula of accuracy is

\begin{equation}
\eta=\frac{n_{right}}{n_{exe}},
\end{equation}

where $n_{right}$ represents the number of correct discriminations in the test set, and $n_{exe}$ represents the test, the optimizer we choose the algorithm of formula $\mathrm{AdaDelta}$. The running average $E[g^2]_{t}$ at time step $t$ depends (as a fraction $\alpha$ similar to the Momentum term. We set $\alpha$ to a similar value as the momentum term, around $0.9$.) on the previous average and the current gradient.

\begin{equation}
\begin{cases}E[g^2]_{t}=\alpha E[g^2]_{t-1}+(1-\alpha)g_{t}^2,\\[2ex]\Delta W_{t}=-\dfrac{\sqrt{\sum_{i=1}^{t-1}\Delta W_{i}}}{\sqrt{E[g^2]_{t}+\epsilon}},\\[2ex]W_{t+1}=W_{t}+\Delta W_{t},
\end{cases}
\end{equation}

This algorithm can speed up the training through the cross-entropy loss function in the formula.

\begin{equation}
\mathrm{Loss(\hat{x},x)}=-\sum_{i=1}^nx\log\mathrm{(\hat{x})},
\end{equation}

Where $x$ is the real mark of the sample, and $\hat{x}$ is the model prediction probability of the sample. In order to achieve more efficient network training, we choose the adam optimizer to adjust the learning rate of each parameter by calculating the first-order moment estimation and the second-order moment estimation of the gradient.

To analyze the error between the predicted values and labels of different dimensions. As shown in Figures 3(b) and 3(d), the change of $g_{3}$ leads to a more complicated spectrum change. Unsurprisingly, we take more than $5000$ samples to measure the variation of the $g_{3}$ parameter, and reach the accuracy of $99.5\%$ as shown in Figure 7(b). The more complex the spectrum, the more training samples are required. We change the optimizer into a more robust $\mathrm{Adam}$ algorithm, as shown in Figure 5(e), and add a convolution layer to the neural network in order to improve the training efficiency.

Next, we take the same parameters change for measuring the co-variation of $d_{2}$ and $g_{3}$. Each sample contains $100$ data points, and the $100$ data points are labeled from $0$ to $99$. The $1$ digit and the $10$ digit represent the size change of two different parameters, respectively. By circulating the different simulation times, we obtain the different samples. Keeping the same neural network structure when $g_{3}$ is recognized, the multi-layer linear layer can achieve a higher level of accuracy, as shown in Figure 5(e). Adding more convolution activation layers can improve the accuracy and stability. In order to improve the training efficiency, we increase the number of convolution kernels in the convolution activation layer. The accuracy attains to $97.5\%$ through $10000$ training samples as shown in Figure 7(c), which is higher than that in the literature~\cite{li2021smart}. Through our theoretical simulation, we prove that we can distinguish the spectral characteristics of multi-parameter changes by machine learning, and the machine learning has the potential to improve the recognition accuracy. Our theoretical model has the potential to find out all kinds of factors affecting the quantum OFCs spectrum. \\

\textbf{5. Conclusion}\\

In our theoretical research, based on the quantum OFCs's multi-mode characteristics, we firstly change the structural parameters of micro-ring array, and then we use machine learning methods to identify different quantum OFCs spectrum, and finally we achieve a more accuracy spectrum information.In a word, our theoretical work can prove that it is feasible to realize on-chip high precision information detection by using the quantum OFCs spectrum of micro-ring array.

\begin{table}
\caption{\label{table} The change of $d_{2}$ recognition accuracy with the amount of training data.}
\begin{tabular}{|p{40pt}|p{40pt}|p{40pt}|p{40pt}|p{40pt}|p{40pt}|}
\hline
param & 
4000& 
4125&
4250&
4375&
4500\\
\hline
 $d_{2}$ & 1 & 0.996 &  0.998  &  0.998 & 1 \\
  $d_{2}$ & 0.996 & 0.996 & 0.998  & 0.998   & 1\\
$d_{2}$ & 1&0.998 & 1  & 1 & 1 \\
 $d_{2}$ &1 & 1 & 1  &0.998&0.996 \\
$d_{2}$ &1& 1 & 1 &1&1 \\
\hline
\end{tabular}
\end{table}

\begin{table}
\caption{\label{table}The change of $g_{3}$ recognition accuracy with the amount of data.}
\begin{tabular}{|p{20pt}|p{45pt}|p{45pt}|p{45pt}|p{45pt}|p{45pt}|}
\hline
param & 40000  &  41250  &  42500  &  43750 & 45000 \\
\hline
 $g_{3}$ & 0.98637	&0.99619&	0.99519	&0.99559	&0.99659  \\  
   $g_{3}$ & 0.99238&	0.99359&	0.99599&	0.99499&	0.9982  \\  
   $g_{3}$ & 0.99319	&0.99098&	0.99399	&0.99639	&0.99499 \\  
   $g_{3}$ &0.99279	&0.99539	&0.99198	&0.96954&	0.99279 \\  
   $g_{3}$ &0.99158	&0.99499	&0.99218&	0.99319	&0.99218 \\  
\hline
\end{tabular}
\end{table}

\begin{table}
\caption{\label{table} The accuracy of $d_{2}$ and $g_{3}$ parameter identification varies with the amount of data.}
\begin{tabular}{|p{40pt}|p{45pt}|p{45pt}|p{45pt}|p{45pt}|p{45pt}|}
\hline
param & 
880000 & 
 900000 &
920000 &
940000 &
960000 \\
\hline
$d_{2}$,$g_{3}$ & 0.96828	&0.95862	&0.9679	&0.96862	&0.97194 \\  
   $d_{2}$,$g_{3}$ &0.96932&	0.96664&	0.97272&	0.9654	&0.9675  \\  
   $d_{2}$,$g_{3}$ &0.97166&	0.96732&	0.96552	&0.97168	&0.96836 \\  
   $d_{2}$,$g_{3}$ &0.97036	&0.96864&	0.96096&	0.97118&	0.93958  \\  
   $d_{2}$,$g_{3}$ &0.9632&	0.96172	&0.95812	&0.95684	&0.96976  \\  
\hline
\end{tabular}
\end{table}

\textbf{Funding.}
Financial support from the project funded by State Key Laboratory of Quantum Optics Technologies and Devices, Shanxi University, Taiyuan, China (Grants No.KF202004, KF202205, KF202503).\\

\textbf{Disclosures.} The authors declare no conflicts of interest.\\

\textbf{Data availability statement.} Data underlying the results presented in this paper are not publicly available at this time but may be obtained from the authors upon reasonable request.\\

\bibliography{jsen}

\begin{thebibliography}{27}%
\makeatletter
\providecommand \@ifxundefined [1]{%
 \@ifx{#1\undefined}
}%
\providecommand \@ifnum [1]{%
 \ifnum #1\expandafter \@firstoftwo
 \else \expandafter \@secondoftwo
 \fi
}%
\providecommand \@ifx [1]{%
 \ifx #1\expandafter \@firstoftwo
 \else \expandafter \@secondoftwo
 \fi
}%
\providecommand \natexlab [1]{#1}%
\providecommand \enquote  [1]{``#1''}%
\providecommand \bibnamefont  [1]{#1}%
\providecommand \bibfnamefont [1]{#1}%
\providecommand \citenamefont [1]{#1}%
\providecommand \href@noop [0]{\@secondoftwo}%
\providecommand \href [0]{\begingroup \@sanitize@url \@href}%
\providecommand \@href[1]{\@@startlink{#1}\@@href}%
\providecommand \@@href[1]{\endgroup#1\@@endlink}%
\providecommand \@sanitize@url [0]{\catcode `\\12\catcode `\$12\catcode `\&12\catcode `\#12\catcode `\^12\catcode `\_12\catcode `\%12\relax}%
\providecommand \@@startlink[1]{}%
\providecommand \@@endlink[0]{}%
\providecommand \url  [0]{\begingroup\@sanitize@url \@url }%
\providecommand \@url [1]{\endgroup\@href {#1}{\urlprefix }}%
\providecommand \urlprefix  [0]{URL }%
\providecommand \Eprint [0]{\href }%
\providecommand \doibase [0]{https://doi.org/}%
\providecommand \selectlanguage [0]{\@gobble}%
\providecommand \bibinfo  [0]{\@secondoftwo}%
\providecommand \bibfield  [0]{\@secondoftwo}%
\providecommand \translation [1]{[#1]}%
\providecommand \BibitemOpen [0]{}%
\providecommand \bibitemStop [0]{}%
\providecommand \bibitemNoStop [0]{.\EOS\space}%
\providecommand \EOS [0]{\spacefactor3000\relax}%
\providecommand \BibitemShut  [1]{\csname bibitem#1\endcsname}%
\let\auto@bib@innerbib\@empty
\bibitem [{\citenamefont {Jiang}\ \emph {et~al.}(2020)\citenamefont {Jiang}, \citenamefont {Qavi}, \citenamefont {Huang},\ and\ \citenamefont {Yang}}]{jiang2020whispering}%
  \BibitemOpen
  \bibfield  {author} {\bibinfo {author} {\bibfnamefont {X.}~\bibnamefont {Jiang}}, \bibinfo {author} {\bibfnamefont {A.~J.}\ \bibnamefont {Qavi}}, \bibinfo {author} {\bibfnamefont {S.~H.}\ \bibnamefont {Huang}},\ and\ \bibinfo {author} {\bibfnamefont {L.}~\bibnamefont {Yang}},\ }\bibfield  {title} {\bibinfo {title} {Whispering-gallery sensors},\ }\href@noop {} {\bibfield  {journal} {\bibinfo  {journal} {Matter}\ }\textbf {\bibinfo {volume} {3}},\ \bibinfo {pages} {371} (\bibinfo {year} {2020})}\BibitemShut {NoStop}%
\bibitem [{\citenamefont {Xue}\ \emph {et~al.}(2019)\citenamefont {Xue}, \citenamefont {Liang}, \citenamefont {Li}, \citenamefont {Sun}, \citenamefont {Xiang}, \citenamefont {Zhang}, \citenamefont {Dai}, \citenamefont {Duo}, \citenamefont {Wu}, \citenamefont {Qi} \emph {et~al.}}]{xue2019ultrasensitive}%
  \BibitemOpen
  \bibfield  {author} {\bibinfo {author} {\bibfnamefont {T.}~\bibnamefont {Xue}}, \bibinfo {author} {\bibfnamefont {W.}~\bibnamefont {Liang}}, \bibinfo {author} {\bibfnamefont {Y.}~\bibnamefont {Li}}, \bibinfo {author} {\bibfnamefont {Y.}~\bibnamefont {Sun}}, \bibinfo {author} {\bibfnamefont {Y.}~\bibnamefont {Xiang}}, \bibinfo {author} {\bibfnamefont {Y.}~\bibnamefont {Zhang}}, \bibinfo {author} {\bibfnamefont {Z.}~\bibnamefont {Dai}}, \bibinfo {author} {\bibfnamefont {Y.}~\bibnamefont {Duo}}, \bibinfo {author} {\bibfnamefont {L.}~\bibnamefont {Wu}}, \bibinfo {author} {\bibfnamefont {K.}~\bibnamefont {Qi}}, \emph {et~al.},\ }\bibfield  {title} {\bibinfo {title} {Ultrasensitive detection of mirna with an antimonene-based surface plasmon resonance sensor},\ }\href@noop {} {\bibfield  {journal} {\bibinfo  {journal} {Nature communications}\ }\textbf {\bibinfo {volume} {10}},\ \bibinfo {pages} {28} (\bibinfo {year} {2019})}\BibitemShut {NoStop}%
\bibitem [{\citenamefont {Kuhnline}\ \emph {et~al.}(2013)\citenamefont {Kuhnline} \emph {et~al.}}]{kuhnline2013interfacing}%
  \BibitemOpen
  \bibfield  {author} {\bibinfo {author} {\bibfnamefont {S.~C.~D.}\ \bibnamefont {Kuhnline}} \emph {et~al.},\ }\bibfield  {title} {\bibinfo {title} {Interfacing lipid bilayer nanodiscs and silicon photonic sensor arrays for multiplexed protein--lipid and protein--membrane protein interaction screening},\ }\href@noop {} {\bibfield  {journal} {\bibinfo  {journal} {Analytical Chemistry}\ }\textbf {\bibinfo {volume} {85}},\ \bibinfo {pages} {2970} (\bibinfo {year} {2013})}\BibitemShut {NoStop}%
\bibitem [{\citenamefont {Chen}\ \emph {et~al.}(2023)\citenamefont {Chen}, \citenamefont {Chen}, \citenamefont {Fu}, \citenamefont {Xie}, \citenamefont {Lu},\ and\ \citenamefont {Zhang}}]{chen2023optical}%
  \BibitemOpen
  \bibfield  {author} {\bibinfo {author} {\bibfnamefont {Q.}~\bibnamefont {Chen}}, \bibinfo {author} {\bibfnamefont {L.}~\bibnamefont {Chen}}, \bibinfo {author} {\bibfnamefont {Z.}~\bibnamefont {Fu}}, \bibinfo {author} {\bibfnamefont {S.}~\bibnamefont {Xie}}, \bibinfo {author} {\bibfnamefont {Q.}~\bibnamefont {Lu}},\ and\ \bibinfo {author} {\bibfnamefont {X.}~\bibnamefont {Zhang}},\ }\bibfield  {title} {\bibinfo {title} {Optical frequency comb-based aerostatic micro pressure sensor aided by machine learning},\ }\href@noop {} {\bibfield  {journal} {\bibinfo  {journal} {IEEE Sensors Journal}\ }\textbf {\bibinfo {volume} {23}},\ \bibinfo {pages} {21078} (\bibinfo {year} {2023})}\BibitemShut {NoStop}%
\bibitem [{\citenamefont {Wade}\ \emph {et~al.}(2015)\citenamefont {Wade}, \citenamefont {Alsop}, \citenamefont {Vertin}, \citenamefont {Yang}, \citenamefont {Johnson},\ and\ \citenamefont {Bailey}}]{wade2015rapid}%
  \BibitemOpen
  \bibfield  {author} {\bibinfo {author} {\bibfnamefont {J.~H.}\ \bibnamefont {Wade}}, \bibinfo {author} {\bibfnamefont {A.~T.}\ \bibnamefont {Alsop}}, \bibinfo {author} {\bibfnamefont {N.~R.}\ \bibnamefont {Vertin}}, \bibinfo {author} {\bibfnamefont {H.}~\bibnamefont {Yang}}, \bibinfo {author} {\bibfnamefont {M.~D.}\ \bibnamefont {Johnson}},\ and\ \bibinfo {author} {\bibfnamefont {R.~C.}\ \bibnamefont {Bailey}},\ }\bibfield  {title} {\bibinfo {title} {Rapid, multiplexed phosphoprotein profiling using silicon photonic sensor arrays},\ }\href@noop {} {\bibfield  {journal} {\bibinfo  {journal} {ACS Central Science}\ }\textbf {\bibinfo {volume} {1}},\ \bibinfo {pages} {374} (\bibinfo {year} {2015})}\BibitemShut {NoStop}%
\bibitem [{\citenamefont {Gohring}\ \emph {et~al.}(2010)\citenamefont {Gohring}, \citenamefont {Dale},\ and\ \citenamefont {Fan}}]{gohring2010detection}%
  \BibitemOpen
  \bibfield  {author} {\bibinfo {author} {\bibfnamefont {J.~T.}\ \bibnamefont {Gohring}}, \bibinfo {author} {\bibfnamefont {P.~S.}\ \bibnamefont {Dale}},\ and\ \bibinfo {author} {\bibfnamefont {X.}~\bibnamefont {Fan}},\ }\bibfield  {title} {\bibinfo {title} {Detection of her2 breast cancer biomarker using the opto-fluidic ring resonator biosensor},\ }\href@noop {} {\bibfield  {journal} {\bibinfo  {journal} {Sensors and Actuators B: Chemical}\ }\textbf {\bibinfo {volume} {146}},\ \bibinfo {pages} {226} (\bibinfo {year} {2010})}\BibitemShut {NoStop}%
\bibitem [{\citenamefont {Li}\ \emph {et~al.}(2021)\citenamefont {Li}, \citenamefont {Zhang}, \citenamefont {Nguyen}, \citenamefont {Luo}, \citenamefont {Liu}, \citenamefont {Zou}, \citenamefont {Shi}, \citenamefont {Cai}, \citenamefont {Yang}, \citenamefont {Jin} \emph {et~al.}}]{li2021smart}%
  \BibitemOpen
  \bibfield  {author} {\bibinfo {author} {\bibfnamefont {Z.}~\bibnamefont {Li}}, \bibinfo {author} {\bibfnamefont {H.}~\bibnamefont {Zhang}}, \bibinfo {author} {\bibfnamefont {B.~T.~T.}\ \bibnamefont {Nguyen}}, \bibinfo {author} {\bibfnamefont {S.}~\bibnamefont {Luo}}, \bibinfo {author} {\bibfnamefont {P.~Y.}\ \bibnamefont {Liu}}, \bibinfo {author} {\bibfnamefont {J.}~\bibnamefont {Zou}}, \bibinfo {author} {\bibfnamefont {Y.}~\bibnamefont {Shi}}, \bibinfo {author} {\bibfnamefont {H.}~\bibnamefont {Cai}}, \bibinfo {author} {\bibfnamefont {Z.}~\bibnamefont {Yang}}, \bibinfo {author} {\bibfnamefont {Y.}~\bibnamefont {Jin}}, \emph {et~al.},\ }\bibfield  {title} {\bibinfo {title} {Smart ring resonator-based sensor for multicomponent chemical analysis via machine learning},\ }\href@noop {} {\bibfield  {journal} {\bibinfo  {journal} {Photonics Research}\ }\textbf {\bibinfo {volume} {9}},\ \bibinfo {pages} {B38} (\bibinfo {year} {2021})}\BibitemShut {NoStop}%
\bibitem [{\citenamefont {Wade}\ and\ \citenamefont {Bailey}(2016)}]{wade2016applications}%
  \BibitemOpen
  \bibfield  {author} {\bibinfo {author} {\bibfnamefont {J.~H.}\ \bibnamefont {Wade}}\ and\ \bibinfo {author} {\bibfnamefont {R.~C.}\ \bibnamefont {Bailey}},\ }\bibfield  {title} {\bibinfo {title} {Applications of optical microcavity resonators in analytical chemistry},\ }\href@noop {} {\bibfield  {journal} {\bibinfo  {journal} {Annual Review of Analytical Chemistry}\ }\textbf {\bibinfo {volume} {9}},\ \bibinfo {pages} {1} (\bibinfo {year} {2016})}\BibitemShut {NoStop}%
\bibitem [{\citenamefont {Sun}\ and\ \citenamefont {Fan}(2011)}]{sun2011optical}%
  \BibitemOpen
  \bibfield  {author} {\bibinfo {author} {\bibfnamefont {Y.}~\bibnamefont {Sun}}\ and\ \bibinfo {author} {\bibfnamefont {X.}~\bibnamefont {Fan}},\ }\bibfield  {title} {\bibinfo {title} {Optical ring resonators for biochemical and chemical sensing},\ }\href@noop {} {\bibfield  {journal} {\bibinfo  {journal} {Analytical and bioanalytical chemistry}\ }\textbf {\bibinfo {volume} {399}},\ \bibinfo {pages} {205} (\bibinfo {year} {2011})}\BibitemShut {NoStop}%
\bibitem [{\citenamefont {Nath}\ \emph {et~al.}(2007)\citenamefont {Nath}, \citenamefont {Atkins},\ and\ \citenamefont {Sligar}}]{nath2007applications}%
  \BibitemOpen
  \bibfield  {author} {\bibinfo {author} {\bibfnamefont {A.}~\bibnamefont {Nath}}, \bibinfo {author} {\bibfnamefont {W.~M.}\ \bibnamefont {Atkins}},\ and\ \bibinfo {author} {\bibfnamefont {S.~G.}\ \bibnamefont {Sligar}},\ }\bibfield  {title} {\bibinfo {title} {Applications of phospholipid bilayer nanodiscs in the study of membranes and membrane proteins},\ }\href@noop {} {\bibfield  {journal} {\bibinfo  {journal} {Biochemistry}\ }\textbf {\bibinfo {volume} {46}},\ \bibinfo {pages} {2059} (\bibinfo {year} {2007})}\BibitemShut {NoStop}%
\bibitem [{\citenamefont {Wehrens}\ and\ \citenamefont {Mevik}(2007)}]{wehrens2007pls}%
  \BibitemOpen
  \bibfield  {author} {\bibinfo {author} {\bibfnamefont {R.}~\bibnamefont {Wehrens}}\ and\ \bibinfo {author} {\bibfnamefont {B.-H.}\ \bibnamefont {Mevik}},\ }\bibfield  {title} {\bibinfo {title} {The pls package: principal component and partial least squares regression in r},\ }\href@noop {} {\bibfield  {journal} {\bibinfo  {journal} {Journal of Statistical Software}\ }\textbf {\bibinfo {volume} {18}},\ \bibinfo {pages} {1} (\bibinfo {year} {2007})}\BibitemShut {NoStop}%
\bibitem [{\citenamefont {Roggo}\ \emph {et~al.}(2007)\citenamefont {Roggo}, \citenamefont {Chalus}, \citenamefont {Maurer}, \citenamefont {Lema-Martinez}, \citenamefont {Edmond},\ and\ \citenamefont {Jent}}]{roggo2007review}%
  \BibitemOpen
  \bibfield  {author} {\bibinfo {author} {\bibfnamefont {Y.}~\bibnamefont {Roggo}}, \bibinfo {author} {\bibfnamefont {P.}~\bibnamefont {Chalus}}, \bibinfo {author} {\bibfnamefont {L.}~\bibnamefont {Maurer}}, \bibinfo {author} {\bibfnamefont {C.}~\bibnamefont {Lema-Martinez}}, \bibinfo {author} {\bibfnamefont {A.}~\bibnamefont {Edmond}},\ and\ \bibinfo {author} {\bibfnamefont {N.}~\bibnamefont {Jent}},\ }\bibfield  {title} {\bibinfo {title} {A review of near infrared spectroscopy and chemometrics in pharmaceutical technologies},\ }\href@noop {} {\bibfield  {journal} {\bibinfo  {journal} {Journal of pharmaceutical and biomedical analysis}\ }\textbf {\bibinfo {volume} {44}},\ \bibinfo {pages} {683} (\bibinfo {year} {2007})}\BibitemShut {NoStop}%
\bibitem [{\citenamefont {Toumi}\ \emph {et~al.}(2014)\citenamefont {Toumi}, \citenamefont {Caldarelli},\ and\ \citenamefont {Torrésani}}]{toumi2014review}%
  \BibitemOpen
  \bibfield  {author} {\bibinfo {author} {\bibfnamefont {I.}~\bibnamefont {Toumi}}, \bibinfo {author} {\bibfnamefont {S.}~\bibnamefont {Caldarelli}},\ and\ \bibinfo {author} {\bibfnamefont {B.}~\bibnamefont {Torrésani}},\ }\bibfield  {title} {\bibinfo {title} {A review of blind source separation in nmr spectroscopy},\ }\href@noop {} {\bibfield  {journal} {\bibinfo  {journal} {Progress in nuclear magnetic resonance spectroscopy}\ }\textbf {\bibinfo {volume} {81}},\ \bibinfo {pages} {37} (\bibinfo {year} {2014})}\BibitemShut {NoStop}%
\bibitem [{\citenamefont {LeCun}\ \emph {et~al.}(2015)\citenamefont {LeCun}, \citenamefont {Bengio},\ and\ \citenamefont {Hinton}}]{lecun2015deep}%
  \BibitemOpen
  \bibfield  {author} {\bibinfo {author} {\bibfnamefont {Y.}~\bibnamefont {LeCun}}, \bibinfo {author} {\bibfnamefont {Y.}~\bibnamefont {Bengio}},\ and\ \bibinfo {author} {\bibfnamefont {G.}~\bibnamefont {Hinton}},\ }\bibfield  {title} {\bibinfo {title} {Deep learning},\ }\href@noop {} {\bibfield  {journal} {\bibinfo  {journal} {nature}\ }\textbf {\bibinfo {volume} {521}},\ \bibinfo {pages} {436} (\bibinfo {year} {2015})}\BibitemShut {NoStop}%
\bibitem [{\citenamefont {Schmidhuber}(2015)}]{schmidhuber2015deep}%
  \BibitemOpen
  \bibfield  {author} {\bibinfo {author} {\bibfnamefont {J.}~\bibnamefont {Schmidhuber}},\ }\bibfield  {title} {\bibinfo {title} {Deep learning in neural networks: An overview},\ }\href@noop {} {\bibfield  {journal} {\bibinfo  {journal} {Neural networks}\ }\textbf {\bibinfo {volume} {61}},\ \bibinfo {pages} {85} (\bibinfo {year} {2015})}\BibitemShut {NoStop}%
\bibitem [{\citenamefont {Robison}\ \emph {et~al.}(2019)\citenamefont {Robison}, \citenamefont {Escalante}, \citenamefont {Valera}, \citenamefont {Erskine}, \citenamefont {Auvil}, \citenamefont {Sasieta}, \citenamefont {Bushell}, \citenamefont {Welge},\ and\ \citenamefont {Bailey}}]{robison2019precision}%
  \BibitemOpen
  \bibfield  {author} {\bibinfo {author} {\bibfnamefont {H.~M.}\ \bibnamefont {Robison}}, \bibinfo {author} {\bibfnamefont {P.}~\bibnamefont {Escalante}}, \bibinfo {author} {\bibfnamefont {E.}~\bibnamefont {Valera}}, \bibinfo {author} {\bibfnamefont {C.~L.}\ \bibnamefont {Erskine}}, \bibinfo {author} {\bibfnamefont {L.}~\bibnamefont {Auvil}}, \bibinfo {author} {\bibfnamefont {H.~C.}\ \bibnamefont {Sasieta}}, \bibinfo {author} {\bibfnamefont {C.}~\bibnamefont {Bushell}}, \bibinfo {author} {\bibfnamefont {M.}~\bibnamefont {Welge}},\ and\ \bibinfo {author} {\bibfnamefont {R.~C.}\ \bibnamefont {Bailey}},\ }\bibfield  {title} {\bibinfo {title} {Precision immunoprofiling to reveal diagnostic signatures for latent tuberculosis infection and reactivation risk stratification},\ }\href@noop {} {\bibfield  {journal} {\bibinfo  {journal} {Integrative Biology}\ }\textbf {\bibinfo {volume} {11}},\ \bibinfo {pages} {16} (\bibinfo {year} {2019})}\BibitemShut {NoStop}%
\bibitem [{\citenamefont {Vamathevan}\ \emph {et~al.}(2019)\citenamefont {Vamathevan}, \citenamefont {Clark}, \citenamefont {Czodrowski}, \citenamefont {Dunham}, \citenamefont {Ferran}, \citenamefont {Lee}, \citenamefont {Li}, \citenamefont {Madabhushi}, \citenamefont {Shah}, \citenamefont {Spitzer} \emph {et~al.}}]{vamathevan2019applications}%
  \BibitemOpen
  \bibfield  {author} {\bibinfo {author} {\bibfnamefont {J.}~\bibnamefont {Vamathevan}}, \bibinfo {author} {\bibfnamefont {D.}~\bibnamefont {Clark}}, \bibinfo {author} {\bibfnamefont {P.}~\bibnamefont {Czodrowski}}, \bibinfo {author} {\bibfnamefont {I.}~\bibnamefont {Dunham}}, \bibinfo {author} {\bibfnamefont {E.}~\bibnamefont {Ferran}}, \bibinfo {author} {\bibfnamefont {G.}~\bibnamefont {Lee}}, \bibinfo {author} {\bibfnamefont {B.}~\bibnamefont {Li}}, \bibinfo {author} {\bibfnamefont {A.}~\bibnamefont {Madabhushi}}, \bibinfo {author} {\bibfnamefont {P.}~\bibnamefont {Shah}}, \bibinfo {author} {\bibfnamefont {M.}~\bibnamefont {Spitzer}}, \emph {et~al.},\ }\bibfield  {title} {\bibinfo {title} {Applications of machine learning in drug discovery and development},\ }\href@noop {} {\bibfield  {journal} {\bibinfo  {journal} {Nature reviews Drug discovery}\ }\textbf {\bibinfo {volume} {18}},\ \bibinfo {pages} {463} (\bibinfo {year} {2019})}\BibitemShut {NoStop}%
\bibitem [{\citenamefont {Cheng}\ and\ \citenamefont {Zhao}(2014)}]{cheng2014machine}%
  \BibitemOpen
  \bibfield  {author} {\bibinfo {author} {\bibfnamefont {F.}~\bibnamefont {Cheng}}\ and\ \bibinfo {author} {\bibfnamefont {Z.}~\bibnamefont {Zhao}},\ }\bibfield  {title} {\bibinfo {title} {Machine learning-based prediction of drug--drug interactions by integrating drug phenotypic, therapeutic, chemical, and genomic properties},\ }\href@noop {} {\bibfield  {journal} {\bibinfo  {journal} {Journal of the American Medical Informatics Association}\ }\textbf {\bibinfo {volume} {21}},\ \bibinfo {pages} {e278} (\bibinfo {year} {2014})}\BibitemShut {NoStop}%
\bibitem [{\citenamefont {Moraru}\ \emph {et~al.}(2010)\citenamefont {Moraru}, \citenamefont {Pesko}, \citenamefont {Porcius}, \citenamefont {Fortuna},\ and\ \citenamefont {Mladenic}}]{moraru2010using}%
  \BibitemOpen
  \bibfield  {author} {\bibinfo {author} {\bibfnamefont {A.}~\bibnamefont {Moraru}}, \bibinfo {author} {\bibfnamefont {M.}~\bibnamefont {Pesko}}, \bibinfo {author} {\bibfnamefont {M.}~\bibnamefont {Porcius}}, \bibinfo {author} {\bibfnamefont {C.}~\bibnamefont {Fortuna}},\ and\ \bibinfo {author} {\bibfnamefont {D.}~\bibnamefont {Mladenic}},\ }\bibfield  {title} {\bibinfo {title} {Using machine learning on sensor data},\ }\href@noop {} {\bibfield  {journal} {\bibinfo  {journal} {Journal of computing and information technology}\ }\textbf {\bibinfo {volume} {18}},\ \bibinfo {pages} {341} (\bibinfo {year} {2010})}\BibitemShut {NoStop}%
\bibitem [{\citenamefont {Zhao}\ \emph {et~al.}(2008)\citenamefont {Zhao}, \citenamefont {Bhushan}, \citenamefont {Santamaria}, \citenamefont {Simon},\ and\ \citenamefont {Davis}}]{zhao2008machine}%
  \BibitemOpen
  \bibfield  {author} {\bibinfo {author} {\bibfnamefont {W.}~\bibnamefont {Zhao}}, \bibinfo {author} {\bibfnamefont {A.}~\bibnamefont {Bhushan}}, \bibinfo {author} {\bibfnamefont {A.~D.}\ \bibnamefont {Santamaria}}, \bibinfo {author} {\bibfnamefont {M.~G.}\ \bibnamefont {Simon}},\ and\ \bibinfo {author} {\bibfnamefont {C.~E.}\ \bibnamefont {Davis}},\ }\bibfield  {title} {\bibinfo {title} {Machine learning: A crucial tool for sensor design},\ }\href@noop {} {\bibfield  {journal} {\bibinfo  {journal} {Algorithms}\ }\textbf {\bibinfo {volume} {1}},\ \bibinfo {pages} {130} (\bibinfo {year} {2008})}\BibitemShut {NoStop}%
\bibitem [{\citenamefont {Shen}\ and\ \citenamefont {Zhao}(2024)}]{shen2023method}%
  \BibitemOpen
  \bibfield  {author} {\bibinfo {author} {\bibfnamefont {H.}~\bibnamefont {Shen}}\ and\ \bibinfo {author} {\bibfnamefont {C.}~\bibnamefont {Zhao}},\ }\bibfield  {title} {\bibinfo {title} {A method for determining the formation position of comb tooth of kerr micro-ring},\ }\href@noop {} {\bibfield  {journal} {\bibinfo  {journal} {IEEE Journal of Quantum Electronics}\ }\textbf {\bibinfo {volume} {60}},\ \bibinfo {pages} {9000107} (\bibinfo {year} {2024})}\BibitemShut {NoStop}%
\bibitem [{\citenamefont {Guidry}\ \emph {et~al.}(2022)\citenamefont {Guidry}, \citenamefont {Lukin}, \citenamefont {Yang}, \citenamefont {Trivedi},\ and\ \citenamefont {Vučković}}]{guidry2022quantum}%
  \BibitemOpen
  \bibfield  {author} {\bibinfo {author} {\bibfnamefont {M.~A.}\ \bibnamefont {Guidry}}, \bibinfo {author} {\bibfnamefont {D.~M.}\ \bibnamefont {Lukin}}, \bibinfo {author} {\bibfnamefont {K.~Y.}\ \bibnamefont {Yang}}, \bibinfo {author} {\bibfnamefont {R.}~\bibnamefont {Trivedi}},\ and\ \bibinfo {author} {\bibfnamefont {J.}~\bibnamefont {Vučković}},\ }\bibfield  {title} {\bibinfo {title} {Quantum optics of soliton microcombs},\ }\href@noop {} {\bibfield  {journal} {\bibinfo  {journal} {Nature Photonics}\ }\textbf {\bibinfo {volume} {16}},\ \bibinfo {pages} {52} (\bibinfo {year} {2022})}\BibitemShut {NoStop}%
\bibitem [{\citenamefont {Guo}\ \emph {et~al.}(2016{\natexlab{a}})\citenamefont {Guo}, \citenamefont {Zou},\ and\ \citenamefont {Tang}}]{guo2016second}%
  \BibitemOpen
  \bibfield  {author} {\bibinfo {author} {\bibfnamefont {X.}~\bibnamefont {Guo}}, \bibinfo {author} {\bibfnamefont {C.-L.}\ \bibnamefont {Zou}},\ and\ \bibinfo {author} {\bibfnamefont {H.~X.}\ \bibnamefont {Tang}},\ }\bibfield  {title} {\bibinfo {title} {Second-harmonic generation in aluminum nitride microrings with 2500\%/w conversion efficiency},\ }\href@noop {} {\bibfield  {journal} {\bibinfo  {journal} {Optica}\ }\textbf {\bibinfo {volume} {3}},\ \bibinfo {pages} {1126} (\bibinfo {year} {2016}{\natexlab{a}})}\BibitemShut {NoStop}%
\bibitem [{\citenamefont {Bruch}\ \emph {et~al.}(2021)\citenamefont {Bruch}, \citenamefont {Liu}, \citenamefont {Gong}, \citenamefont {Surya}, \citenamefont {Li}, \citenamefont {Zou},\ and\ \citenamefont {Tang}}]{bruch2021pockels}%
  \BibitemOpen
  \bibfield  {author} {\bibinfo {author} {\bibfnamefont {A.~W.}\ \bibnamefont {Bruch}}, \bibinfo {author} {\bibfnamefont {X.}~\bibnamefont {Liu}}, \bibinfo {author} {\bibfnamefont {Z.}~\bibnamefont {Gong}}, \bibinfo {author} {\bibfnamefont {J.~B.}\ \bibnamefont {Surya}}, \bibinfo {author} {\bibfnamefont {M.}~\bibnamefont {Li}}, \bibinfo {author} {\bibfnamefont {C.-L.}\ \bibnamefont {Zou}},\ and\ \bibinfo {author} {\bibfnamefont {H.~X.}\ \bibnamefont {Tang}},\ }\bibfield  {title} {\bibinfo {title} {Pockels soliton microcomb},\ }\href@noop {} {\bibfield  {journal} {\bibinfo  {journal} {Nature Photonics}\ }\textbf {\bibinfo {volume} {15}},\ \bibinfo {pages} {21} (\bibinfo {year} {2021})}\BibitemShut {NoStop}%
\bibitem [{\citenamefont {Hansson}\ \emph {et~al.}(2014)\citenamefont {Hansson}, \citenamefont {Modotto},\ and\ \citenamefont {Wabnitz}}]{hansson2014numerical}%
  \BibitemOpen
  \bibfield  {author} {\bibinfo {author} {\bibfnamefont {T.}~\bibnamefont {Hansson}}, \bibinfo {author} {\bibfnamefont {D.}~\bibnamefont {Modotto}},\ and\ \bibinfo {author} {\bibfnamefont {S.}~\bibnamefont {Wabnitz}},\ }\bibfield  {title} {\bibinfo {title} {On the numerical simulation of kerr frequency combs using coupled mode equations},\ }\href@noop {} {\bibfield  {journal} {\bibinfo  {journal} {Optics Communications}\ }\textbf {\bibinfo {volume} {312}},\ \bibinfo {pages} {134} (\bibinfo {year} {2014})}\BibitemShut {NoStop}%
\bibitem [{\citenamefont {Chembo}\ and\ \citenamefont {Yu}(2010)}]{chembo2010modal}%
  \BibitemOpen
  \bibfield  {author} {\bibinfo {author} {\bibfnamefont {Y.~K.}\ \bibnamefont {Chembo}}\ and\ \bibinfo {author} {\bibfnamefont {N.}~\bibnamefont {Yu}},\ }\bibfield  {title} {\bibinfo {title} {Modal expansion approach to optical-frequency-comb generation with monolithic whispering-gallery-mode resonators},\ }\href@noop {} {\bibfield  {journal} {\bibinfo  {journal} {Physical Review A}\ }\textbf {\bibinfo {volume} {82}},\ \bibinfo {pages} {033801} (\bibinfo {year} {2010})}\BibitemShut {NoStop}%
\bibitem [{\citenamefont {Guo}\ \emph {et~al.}(2016{\natexlab{b}})\citenamefont {Guo}, \citenamefont {Zou},\ and\ \citenamefont {Tang}}]{guo201670}%
  \BibitemOpen
  \bibfield  {author} {\bibinfo {author} {\bibfnamefont {X.}~\bibnamefont {Guo}}, \bibinfo {author} {\bibfnamefont {C.-L.}\ \bibnamefont {Zou}},\ and\ \bibinfo {author} {\bibfnamefont {H.~X.}\ \bibnamefont {Tang}},\ }\bibfield  {title} {\bibinfo {title} {70 db long-pass filter on a nanophotonic chip},\ }\href@noop {} {\bibfield  {journal} {\bibinfo  {journal} {Optics express}\ }\textbf {\bibinfo {volume} {24}},\ \bibinfo {pages} {21167} (\bibinfo {year} {2016}{\natexlab{b}})}\BibitemShut {NoStop}%
\end{thebibliography}%
\balance
\end{document}